\documentclass[a4paper,dvips 2t,superscriptaddress,showkeys]{revtex4} 
\usepackage{epsfig,graphics,graphicx,amsmath,amssymb}
\usepackage{graphics}
\usepackage{subfigure}
\usepackage{amssymb}
\usepackage{color}
\usepackage{amsmath}
\usepackage{amsthm}
\usepackage{amsopn}
\def\uno{\mbox{1 \kern-.59em {\rm l}}}

\def\be{\begin{equation}}
\def\ee{\end{equation}}
\def\bea{\begin{eqnarray}}
\def\eea{\end{eqnarray}}

\begin{document}
\title{\bf Quantum coherence in neutrino oscillation in matter}

\author{ Z. Askaripour Ravari}
\affiliation{Department of Physics, Islamic Azad University, North Tehran Branch, Tehran 165115-3311, I.R. Iran}
\author{ M. M. Ettefaghi }\email{mettefaghi@qom.ac.ir}\affiliation{Department of Physics, University of Qom, Ghadir Blvd., Qom 371614-6611, I.R. Iran} 
\author{S. Miraboutalebi}
\affiliation{Department of Physics, Islamic Azad University, North Tehran Branch, Tehran 165115-3311, I.R. Iran}

\begin{abstract}
A closer and more detailed study of neutrino oscillation, in addition to assisting us in founding physics beyond the standard model, can potentially be used to understand the fundamental aspects of quantum mechanics. In particular, we know that the neutrino oscillation occurs because the quantum states of the produced and detected neutrinos are a coherent superposition of the mass eigenstates, and this coherency is maintained during the propagation due to the small mass difference of neutrinos. In this paper, we consider the decoherence due to the neutrino interaction in the material medium with constant density in addition to the decoherence coming from the localization properties. For this purpose, we use $l_1\text{-norm}$ in order to quantify the coherence and investigate its dependence on the matter density. According to our results, in general, the coherence in material medium is less than vacuum. However, there exist exceptions; for some matter densities, the localization coherence lengths become infinite. So, for these cases, $l_1\text{-norm}$ in matter is more than the vacuum.
\end{abstract} 

\maketitle

\section{Introduction}
Neutrino oscillation is one of the most interesting phenomena in quantum mechanics which has been experimentally established \cite{rev.est}. It could provide potentially a suitable room for quantum foundation explorations. The quantum approaches to neutrino oscillation are based on the existence of non-zero and non-degenerate neutrino masses. Although the differences between their masses are non-zero, these differences are so smaller than the energy uncertainty in the creation and detection processes that the neutrino mass eigenstate cannot be distinguished. This point is used in the quantum mechanics approach; the states of the created and detected neutrinos (well-known as flavor eigenstates) are written as a coherent superposition of mass eigenstates \cite{par}. This coherency is preserved for large macroscopic distances. Therefore, using the neutrino oscillation phenomenon, one could study quantum coherence, which is a microscopic quantum effect, in large distances even up to several hundred kilometers away. 

Furthermore, neutrinos interact with the matter inside the medium in which they propagate. It is noticeable that due to the presence of electrons in matter, the matter effect leads to an additional potential energy $V$ ($V=\sqrt{2}G_FN_e(x)$ where $G_F$ and $N_e(x)$ are the Fermi constant and the electron density in the medium, respectively) only for electron neutrinos. In fact, neutrino oscillation is caused by the interference effects of coherent neutrino flavor waves. If the potential $V$ is attributed to the elastic forward scattering of neutrinos off electrons, protons and neutrons within the medium via the electroweak interaction, the coherency of neutrino wave may be preserved and the interference effect does not cease generally. In particular, when the matter density in the medium varies adiabatically with the coordinate $x$, and consequently time $t$ ($x\approx t$), a resonance phenomenon occurs. Thus if the primary neutrino state is the electron neutrino, it is possible an almost complete transition to the other flavors to take place at the resonance point even for small mixing angles. Historically,  Wolfenstein first showed the difference in neutrino oscillation in the material medium  in comparison with the vacuum \cite{wolf}. Later on, Smirnov and Mikheev showed that if the electron density in the matter is such a way that we are at the point of resonance, we have the highest flavor conversion \cite{mikh}. This effect is well-known as the MSW effect, which explains the oscillation of solar neutrinos. 
However, the interaction of neutrinos with the matter can cause the survival and transition probabilities of the electron neutrinos to cease when the density of the matter is much larger than the resonances density. This behavior is nontrivial since corresponding to the two distinguished mass squared differences, there are two values for potential which lead to the resonance for the flavor transition. In this paper, we analyze the oscillation probability of neutrinos propagating in a medium with constant matter density for three flavors and study the decoherence effects on the neutrino oscillation due to the interaction with matter in the medium.


Quantum coherence as a basic ingredient for quantum technologies can be quantified using an appropriate quantum resource theory \cite{plenio,plenio2}. In these frameworks, some meaningful and applicable methods are introduced to quantify the desired quantum resource and provide us with measures in order to determine the quantumness. Among the several measures for quantum coherence,  a very intuitive one is related to the off-diagonal elements of the considered quantum state. Therefore, the $l_1\text{-norm}$ which is defined by the sum of the absolute values of all the off-diagonal elements of the density matrix
\be\label{l1}
\textit{c}(\rho)=\sum_{i\neq j}|\rho_{ij}|,
\ee
is desirable. It has been shown that this quantity can satisfy all conditions of measures in the quantum resource theories \cite{plenio,plenio2}. On the other hand, we know that the neutrino oscillation can take place provided that we have non-zero off-diagonal elements in the corresponding density matrix in the flavor representation basis \cite{glashow}. Hence, the non-zero $l_1\text{-norm}$ or equivalently non-zero coherence is a necessary condition for the neutrino oscillation phenomenon. In this direction, in Ref. \cite{song}, the authors have quantified the coherence
for neutrino oscillation by the $l_1\text{-norm}$ and have compared
it with the experimentally observed data.


Furthermore, in order to have a complete and thorough description of neutrino oscillation, we should use
the wave packet (WP) approach, since the production, propagation
and detection of neutrino should be considered as
localized processes and this localization is very well
fulfilled using the WP approach in neutrino oscillation	\cite{par,kayser,giunti,ettplb,ettscripta,qftrev,matterwp}. In this approach, for a baseline distance more than the coherence length, the oscillation probabilities diminish. Therefore, we 	obtain the coherence length which in this case, depends  not only on the neutrino states properties (mass squared difference,	energy and its dispersion) but also on the matter potential \cite{Holanda,Kersten,smir2021}. Consequently, we obtain the $l_1\text{-norm}$ and analyze its behavior in terms of the baseline distance. In fact, according to its definition by Eq. (\ref{l1}), this parameter depends on all the oscillation probabilities. Therefore, we expect the $l_1\text{-norm}$ is affected nontrivially by the value of $V$.

In the next section, we will extend the theory developed in Ref. \cite{1} for neutrino oscillation in a uniform matter to include the localization properties and analyze the probabilities of oscillation in terms of the baseline distance for some values of the matter potential. The $l_1\text{-norm}$ will be investigated in section \ref{4444}. Finally, we will summarize and discuss on the results in the last section. In Appendix, we will give a brief review of the neutrino oscillation in the uniform matter according to the theory developed in Ref. \cite{1}.

\section{Analyzing oscillation probabilities in matter}\label{3333}
As a proper explanation of the neutrino oscillation phenomenon in the quantum mechanics framework, we describe produced and detected neutrinos by the localized weak interaction processes and suppose that they propagate as a localized WP in the position space \cite{par}. 
Hence, we consider that neutrinos of flavor $\alpha$ are produced at the origin of the space-time coordinates and are detected after a time $T$  by a detector which is located at a distance $L$ and is sensitive to neutrinos of flavor $\beta$.
Furthermore, we suppose that neutrinos propagate in a medium with constant matter density.  In general, their interactions with matter cause the dispersion relation between the momentum $|\mathbf p|$ and the energy $E$ of neutrinos to be changed. Therefore, the investigation of time evolution of the realistic three generations scheme becomes complicated,
generally speaking. Meanwhile, more simplified and accurate arguments have been performed for the oscillations in matter with constant density versus baseline divided by neutrino energy plane by using a perturbative framework in Ref. \cite{1}. We give a brief review of this theory in Appendix A. Accordingly, this theory is based on the effective Hamiltonian matrix which is achievable by using the unjustified either “same energy” or “same momentum” assumption at the plane wave (PW) description level. 
Nevertheless, one can obtain 
the standard formula for the probability of neutrino
oscillations by this assumption if the decoherence effects
due to the WP separation are negligible and
the emitted and absorbed neutrino are sufficiently well
localized \cite{par}. However, the main purpose of this paper is related to considering the decoherence issues. Alternatively, in most situations the
decoherence effects can be reliably
estimated based on the standard oscillation formula and simple physical considerations. 

By assumption a Gaussian wave function for the produced and detected neutrinos, one can obtain the probability of transition in vacuum by using WP approach as follows: \cite{giunti}:
\be
P_{\alpha\beta}(\mathbf{L})\simeq \sum_{i,j}U^*_{\alpha i}U_{\beta i}U_{\alpha j}U^*_{\beta j}\exp\left[-2\pi i\dfrac{L}{L_{ij}^{\text{osc}}}-(\dfrac{L}{L_{ij}^{\text{coh}}})^2-2\pi ^2\rho ^2(\dfrac{\sigma _x}{L_{ij}^{\text{osc}}})^2\right],\label{probability}
\ee
in which $U_{\alpha i}$ denotes the mixing matrix elements. This matrix 
 is unitary and can be parameterized by the well-known {\text{PMNS}} matrix
\be
U_{\text{PMNS}}=\begin{bmatrix}
	c_{13}c_{12} & c_{13}s_{12} &s_{13}e^{-i\delta} \\
	-s_{12}c_{23}-e^{i\delta}c_{12}s_{23}s_{13}&c_{12}c_{23}-e^{i\delta}s_{12}s_{23}s_{13}  &c_{13}s_{23} \\
	s_{12}s_{23}-e^{i\delta}c_{12}c_{23}s_{13}&-c_{12}s_{23}-e^{i\delta}s_{12}c_{23}s_{13} &c_{13}c_{23}  \\
\end{bmatrix}.\label{pmns}
\ee
Here, we use the abbreviation $c_{ij}$ and $s_{ij}$ for $\cos\theta_{ij}$ and $\sin\theta_{ij}$ in which $\theta_{ij}$ stands for the mixing angles.  In Eq. (\ref{probability}), $\sigma_{x}$ is the position uncertainty given by 
\be
\sigma_{x}^{2}\equiv\sigma_{xD}^{2}+\sigma_{xP}^{2},\label{wave packet width}
\ee 
where $\sigma_{xP}$ and $ \sigma_{xD}$ denote the produced and detected neutrino WP widths, respectively, and $\rho$ is a parameter of order unity, related to the energy-momentum conservation of the
production process. Moreover, in this equation, the following definition for $L_{ij}^{\text {osc}}$ and $L_{ij}^{\text{coh}}$ are obtained by direct calculation:
\be
L_{ij}^{\text{osc}}\equiv \dfrac{4\pi E}{\Delta m_{ij}^2},~~~~~~~~~~L_{ij}^{\text{coh}}\equiv \dfrac{4\sqrt{2}\sigma _xE^2}{\mid \Delta m_{ij}^2\mid},
\ee
where the mass square differences are defined by $\Delta m_{ij}^2\equiv m_i^2-m_j^2$. In this manner, we have two more factors in comparison to the PW approach for the probability of oscillation. The first factor is the Gaussian factor which describes the coherence condition. According to this condition, the oscillation can only take place if the overlapping of the mass eigenstates is not destroyed; which means $L < L^{\text{coh}}_{ij}$. The second factor is due to the last term in the exponent which emphasizes that for seeing the oscillation we must have $\sigma_{x}\ll L_{ij}^{\text{osc}}$. 

In the case of propagation in the material medium, the energy eigenvalues and mixing matrix are modified in comparison with the corresponding one in the vacuum. In general, these parameters alter with the position provided that the density of matter depends on it. The WP treatment of the neutrino oscillation in a material medium with variable density is given in Ref. \cite{matterwp}. However, it is complicated to obtain the mixing matrices and energy eigenvalues when we consider the general case of three generations schema. In the case of uniform matter density, these parameters are constant. Therefore, using a treatment similar to Ref. \cite{matterwp}, we can generalize Eq. (\ref{probability}) for the transition probability 
 as follows:
\bea
P^m_{\alpha\beta}(\mathbf{L})&\!\!\simeq\!\! &\sum_{i,j}{U^m}^*_{\alpha i}{U^m}_{\beta i}{U^m}_{\alpha j}{U^m}^*_{\beta j}\nonumber\\ 
&\!\!\times\!\! &\exp\left[-2\pi i\dfrac{L}{{L^m}_{ij}^{\text{osc}}}-(\dfrac{L}{{L^m}_{ij}^{\text{coh}}})^2-2\pi ^2\rho ^2(\dfrac{\sigma _x}{{L^m}_{ij}^{\text{osc}}})^2\right],\label{mprobability}
\eea
in which the superscript $m$ refers to the related parameters in the material medium. ${L^m}_{ij}^{\text{osc}}$ and ${L^m}_{ij}^{\text{coh}}$ are defined in this manner by
\be
{L^m}_{ij}^{\text{osc}}\equiv \dfrac{2\pi}{\Delta E^m_{ij}},~~~~~~~~~~{L^m}_{ij}^{\text{coh}}\equiv \dfrac{2\sqrt{2}\sigma _x}{\mid\Delta v^m_{ij}\mid},\label{mlength}
\ee
where $\Delta E^m_{ij}\equiv E^m_i-E^m_j$ and $\Delta  v^m_{ij}\equiv v^m_i-v^m_j$ in which $E^m_i$'s and $v^m_i$'s are the Hamiltonian eigenvalues and the group velocities in matter, respectively. Here, we suppose that the wavelengths of neutrinos are very small compared to the position uncertainties of the production and detection processes or equivalently, the momenta of neutrinos are very large compared to the corresponding uncertainties. This means that the wave functions of the produced and detected neutrinos in momentum space are very localized around the corresponding averaged momenta. Therefore, we can use approximately the results of Ref. \cite{1} for the mixing matrix and Hamiltonian eigenvalues in Eq. (\ref{mprobability}). 
 In this manner, the mixing matrix $U^m_{\alpha i}$ has the $U_{\text{PMNS}}$ form with differences that $\theta_{12}$ and $\theta_{13}$ are replaced by $\theta^m_{13}$ and $\theta^m_{12}$ which are given, respectively, by
 \be
 \theta_{13}^m=\dfrac{1}{2}\arctan[\dfrac{\epsilon\sin{2\theta_{13}}}{\epsilon\cos{2\theta_{13}}-2EV}],
 \ee
 and 
 \be
 \theta_{12}^m=\dfrac{1}{2}\arctan[\Delta m_{21}^2\dfrac{\sin{2\theta_{12}}}{l_2-l_1}\cos(\theta_{13}-\theta_{13}^m)],
 \ee
 where $\epsilon=\Delta m_{31}^2-\Delta m_{21}^2\sin^2{\theta_{12}}$. Moreover, the Hamiltonian eigenvalues $E^m_i$ are as follows:
 \be
 \label{E12}
 E^m_{1,2}=\frac{1}{4E}[(l_1+l_2)\mp\sqrt{(l_1-l_2)^2+\Delta m_{21}^4\sin^2{2\theta_{12}}\cos^2({\theta_{13}}-\theta_{13}^m})],
 \ee
 and
 \be
 \label{E3}
 E^m_3=\frac{l_3}{2E}.
 \ee
 Here $l_i$'s are given by 
 \be
 l_{1,3}=\dfrac{1}{2}[(\Delta m_{31}^2+2EV+\Delta m^2_{21}\sin^2\theta_{12})\mp\sqrt{(2EV)^2+\epsilon^2-4EV\epsilon \cos{2\theta_{13}}}],\label{l13}
 \ee
 and
 \be
 l_2=\Delta m_{21}^2\cos^2{\theta_{12}}.
 \ee
According to Eq. (\ref{mlength}), in order to obtain the coherence lengths, we need to know the differences of the group velocities of matter eigenstates which are given by
\be
\Delta v^m_{ij}=\frac{d\Delta E^m_{ij}}{dE}.
\ee 
For three generations schema, there exist two independent values for  $\Delta v^m_{ij}$ which are obtained as 
\be
\Delta v^m_{21}=\dfrac{1}{8E}(-8\Delta E_{21}^m+\xi),
\ee
and 
\be
\Delta v^m_{32}=\dfrac{1}{8E}\left(6\frac{d l_3}{dE}-4V-8\Delta E_{32}^m-\dfrac{1}{2}\xi\right),
\ee
where 
\be
\xi=\dfrac{1}{2}\left(\dfrac{1}{4E E_2^m-(l_1+l_2)}\right)\left(8(l_1-l_2)\\\frac{d l_1}{dE}+\dfrac{\zeta}{(\epsilon \cos{2\theta_{13}}-2EV)^2}\right),
\ee
with
\be
\zeta=4\Delta m_{21}^4\epsilon V\cos^2{2\theta_{13}^m}\sin^2{2\theta_{12}}\sin{2\theta_{13}}\sin{2(\theta_{13}-\theta_{13}^m)}.
\ee
Using Eq. (\ref{l13}), one can obtain
\be
\frac{d l_{1,3}}{dE}=V\left(1\mp\dfrac{(2EV-\epsilon\cos{2\theta_{13}})}{\sqrt{\epsilon^2+(2EV)^2-4EV\epsilon\cos{2\theta_{13}}}}
\right).
\ee
For the third the differences of the group velocities, we have $\Delta v^m_{31}=\Delta v^m_{32}+\Delta v^m_{21}$. There are three different values for the density of matter or equivalently for the corresponding potential through which the $\Delta v^m_{ij}$ diminishes and consequently, the related coherence length becomes infinite. 

In fact, it is noticeable that in addition to the dependence on the baseline distance $L$, neutrino energies and the WP widths, the probabilities in medium are function of the matter density through the dependence of $\theta^m_{ij}$ and $\Delta E^m_{ij}$ on it. There exist two resonance points in matter ($V^{\text{res}}_1=3.196\times10^{-16}eV$ and $V^{\text{res}}_2=2.577\times10^{-14}eV$) at which the maximal flavor conversion occurs. The resonances take place provided that the density of matter in the medium is in such a way that the matter mixing angles $\theta_{13}^m$ and $\theta_{12}^m$ become $\frac{\pi}{4}$. We take $E=4.5\times 10^{10}eV$, $\Delta m_{21}^2=7.39\times10^{-5}eV^2$, $\Delta m_{31}^2=2.451\times10^{-3}eV^2$, $\sigma_x=0.5\times 10^{-9}m$, $\theta_{12}=33.82^{\circ}$, $\theta_{13}=8.61^{\circ}$, $\theta_{23}=49.7^{\circ}$ and $\delta=217^{\circ}$ \cite{ie}.
In the following, we analyze the oscillation probabilities for the two following situations:

\begin{itemize}
	\item Let us consider the initially produced neutrinos to be $\nu_e$. We illustrate the survival (diagram a) and transition to $\nu_\mu$ (diagram b) probabilities versus $L$ for various matter densities in Fig. (\ref{f1}).
Explicitly, we plot the probability in terms of $L$ for five cases; vacuum ($V=0$), for three values of potential which correspond to three infinity coherence lengths and for a matter dominated potential. The purple curve corresponds to $V=2.824\times10^{-14} eV$ at which ${L^m}^{\text{coh}}_{32}$ becomes infinite. Since this coherence length does not play significant role in the transition of $\nu_e$, we see that the behavior of survival and transition probabilities is similar to the corresponding situation in vacuum (the green curve) with the difference that damping appears at smaller distance. The red and blue curves depict the survival and transition probabilities given that  $V=2.242 \times10^{-15} eV$ and $V=1.099 \times10^{-14} eV$, respectively. Although with these values of potential, ${L^m}^{\text{coh}}_{21}$, which is relevant for transition $\nu_e$ to $\nu_\mu$, becomes infinite, the amplitudes of oscillation behavior are limited. For the brown curve, we take a matter dominated density ($V=1.00\times10^{-12}eV$). In this case, the transition probability vanishes. 
	\begin{figure}[ht]
	\centering
	\subfigure[]{\includegraphics[scale=0.5]{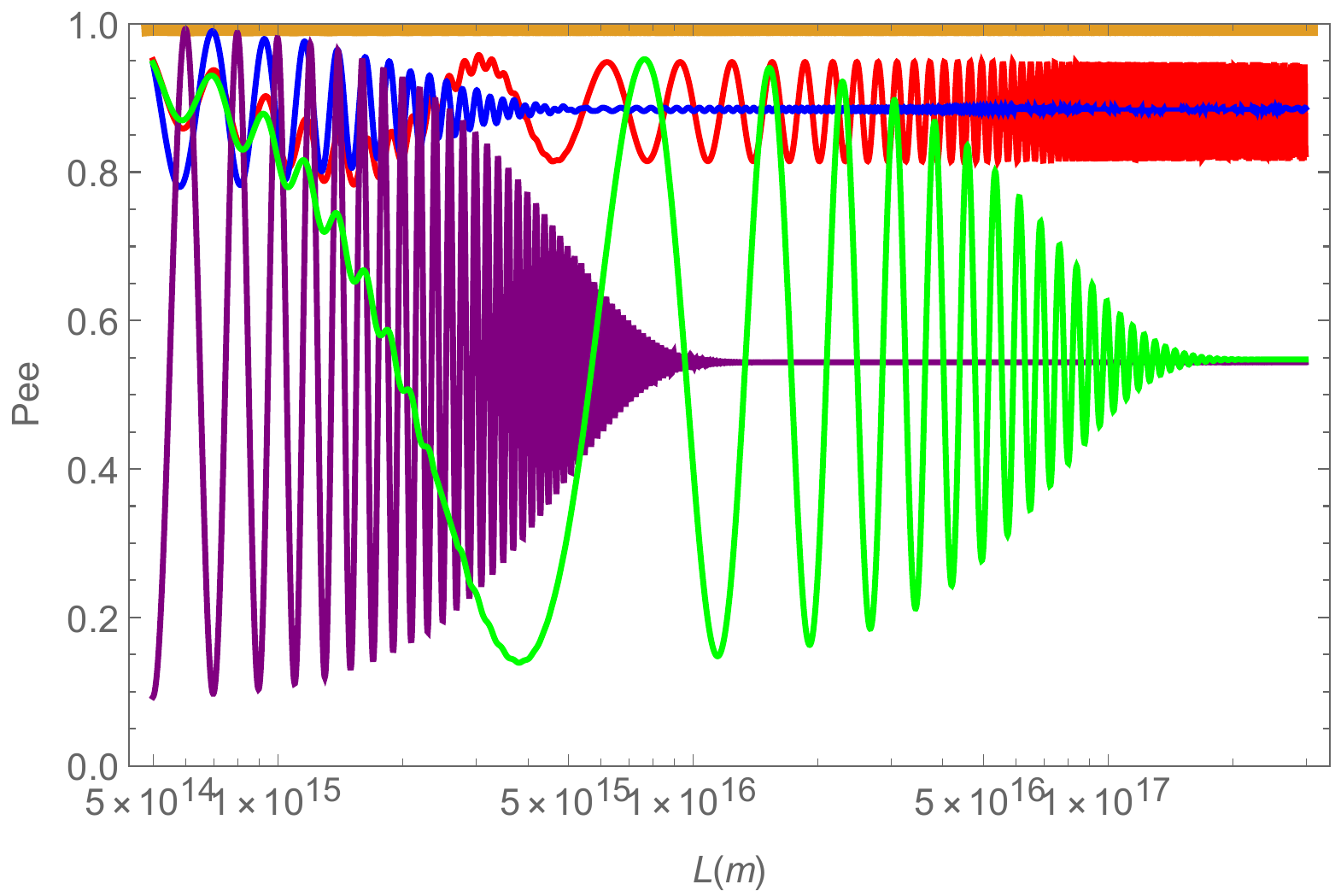}}
	\subfigure[]{\includegraphics[scale=0.5]{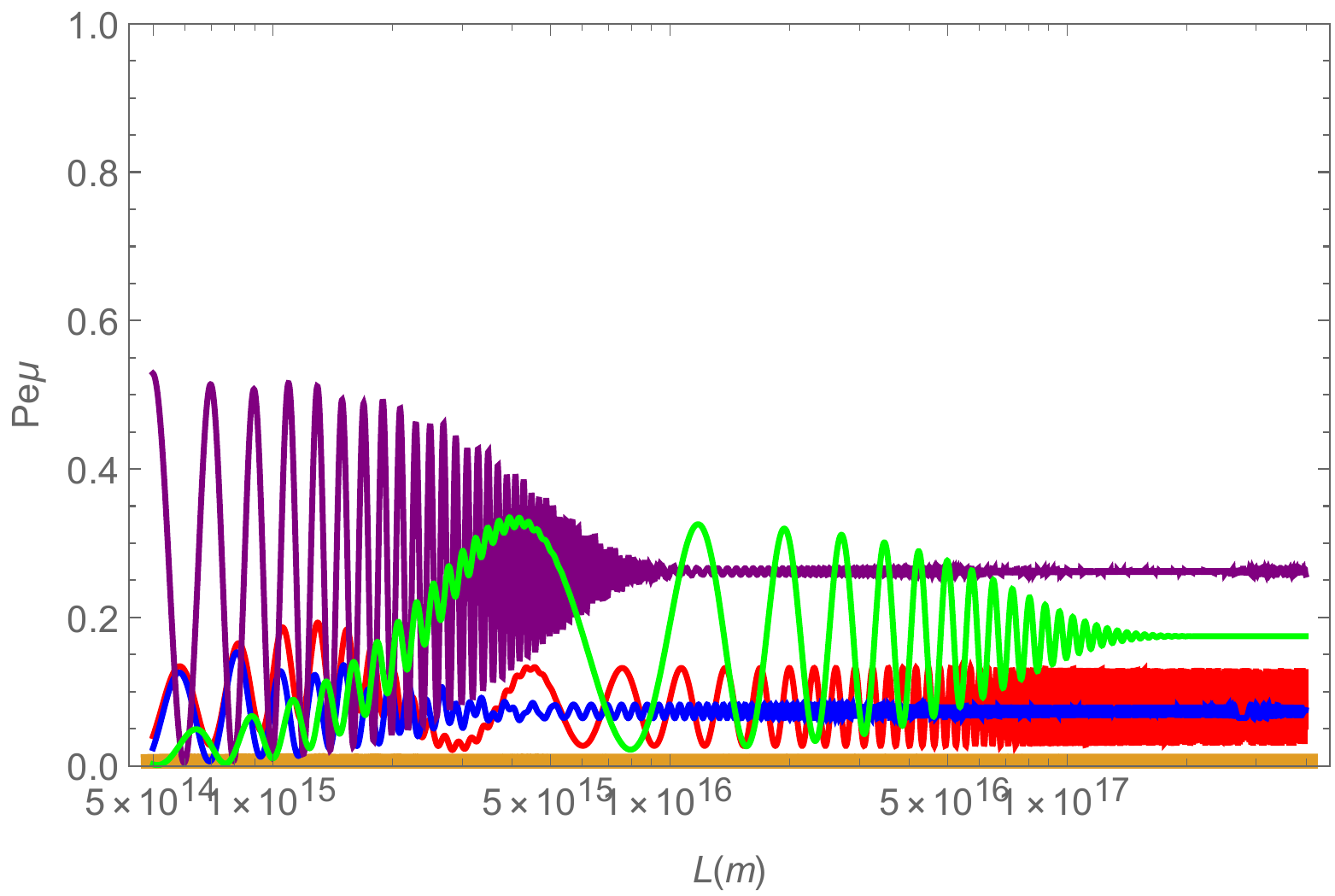}}
	\caption{Survival (a) and transition to $\nu_\mu$ (b) probability of $\nu_e$.  In addition to vacuum (green), we take $V=2.242 \times10^{-15} eV$ (red), $V=1.099 \times10^{-14} eV$(blue) and $V=2.824\times10^{-14} eV$(purple) which correspond to infinity coherence lengths. We also take $V=1.00\times10^{-12} eV$(brown) corresponding to matter dominated.
	}\label{f1}
\end{figure}

\item Now, we consider the initially produced neutrinos to be $\nu_\mu$. We know that the eigenstates $\nu_\mu$ and $\nu_\tau$ interact with matter only through neutral current interactions in contrast to the electron neutrinos. Therefore, it may be imagined that their oscillatory behavior is not affected by the interaction with matter. Hence, we also illustrate the survival (diagram a) and transition to $\nu_e$ (diagram b) probabilities for $\nu_\mu$ versus $L$ for matter densities similar to what were considered in Fig. (\ref{f1}). We see that if $V=2.824\times10^{-14} eV$ in which ${L^m}^{\text{coh}}_{32}$ becomes infinite (the purple curve), the survival probability does not diminish for the assumed interval distance. Meanwhile, the damping behaviors of survival probabilities for the other considered potentials are roughly similar to the vacuum one. In the case of transition to $\nu_e$, we see that the statement is different since ${L^m}^{\text{coh}}_{32}$ is not the relevant coherence length. 


\begin{figure}[ht]
	\centering
	\subfigure[]{\includegraphics[scale=0.5]{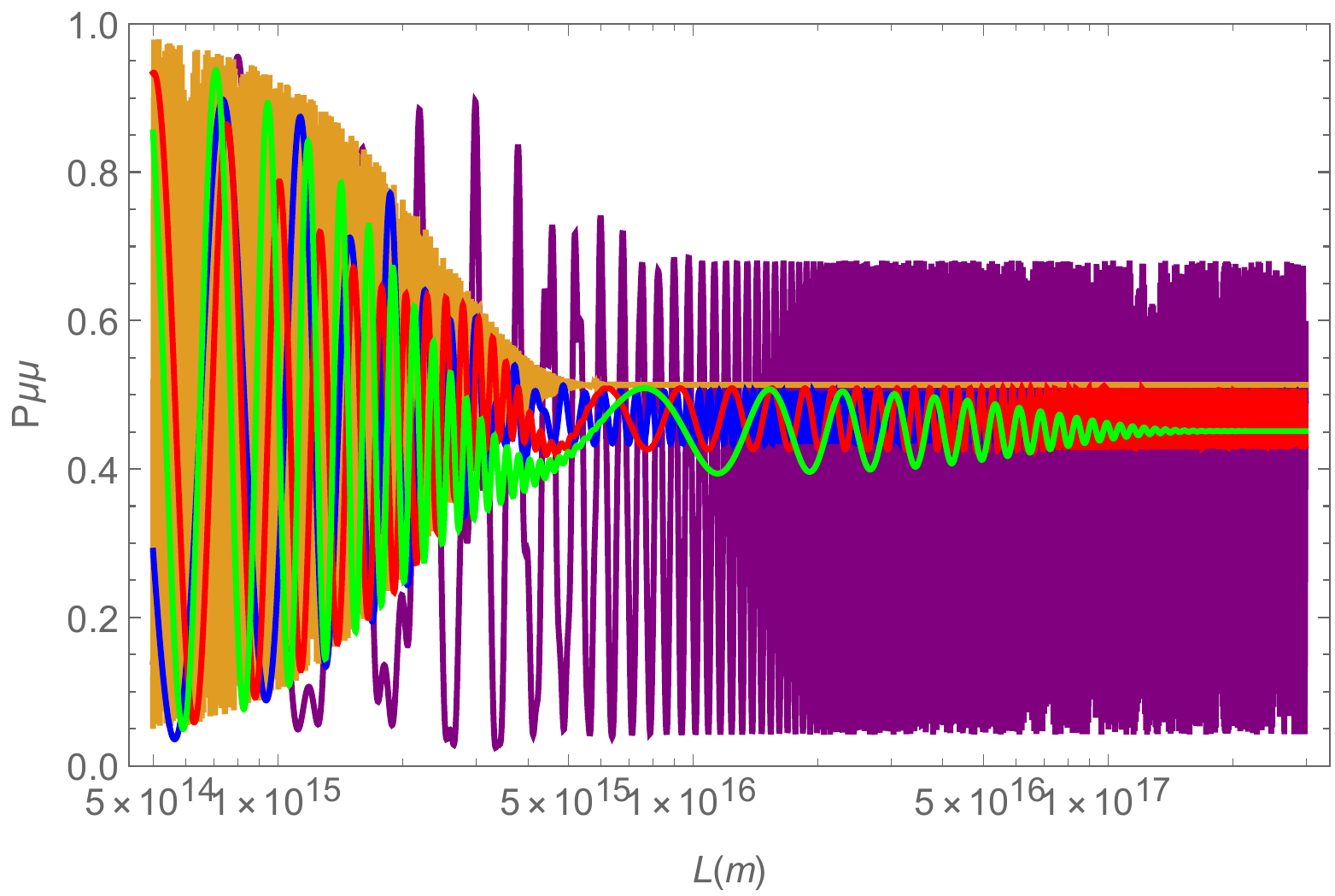}}
	\qquad
\subfigure[]{\includegraphics[scale=0.5]{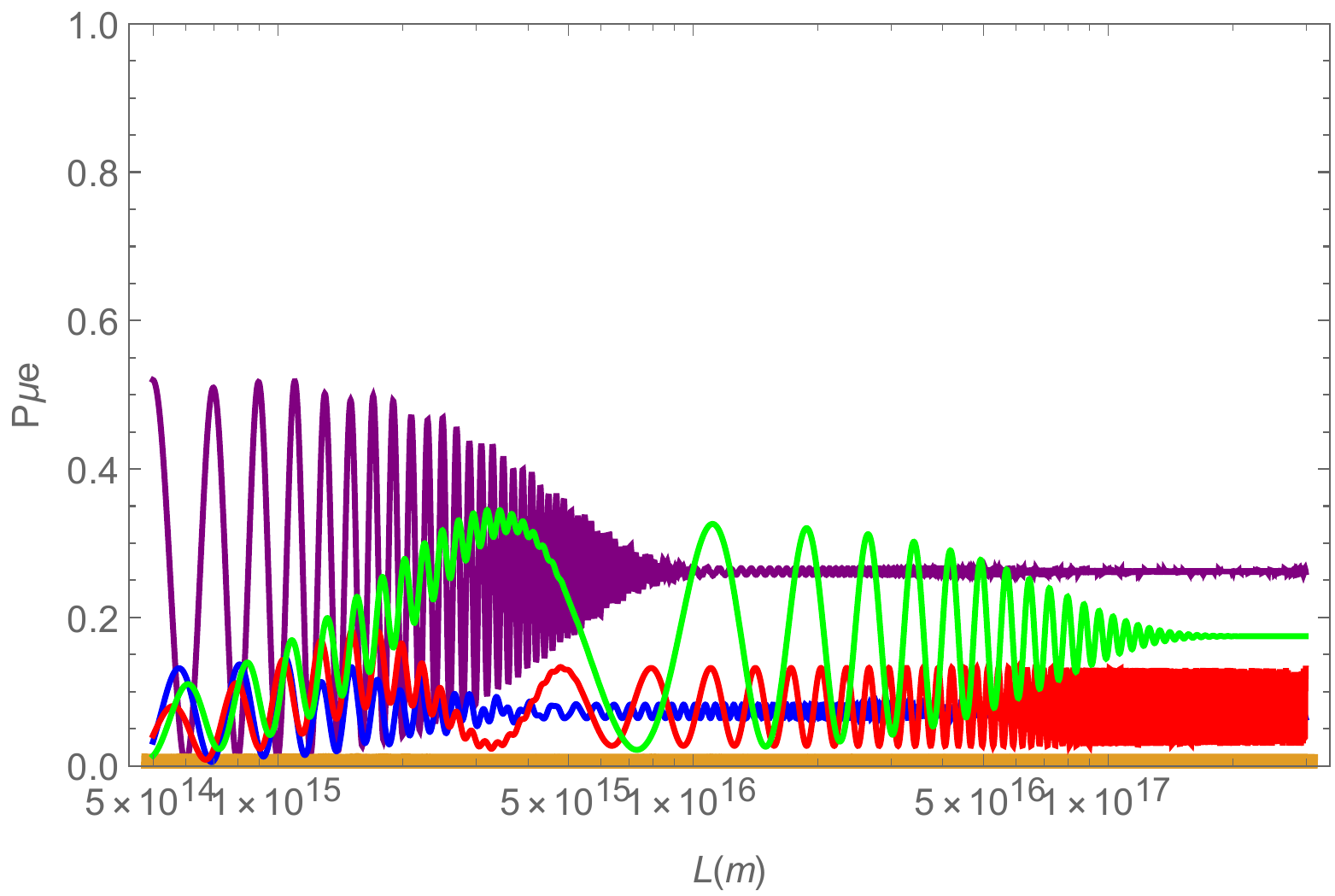}}
\caption{Survival (a) and transition to $\nu_e$ (b) probability of $\nu_\mu$.  In addition to vacuum (green), we take $V=2.242 \times10^{-15} eV$ (red), $V=1.099 \times10^{-14} eV$(blue) and $V=2.824\times10^{-14} eV$(purple) which correspond to infinity coherence lengths. We also take $V=1.00\times10^{-12} eV$(brown) corresponding to matter dominated.}\label{f2}
\end{figure}

\end{itemize}

\section{Study quantum coherence through $l_1\text{-norm}$ calculation}\label{4444}
In order to have the oscillation phenomena for neutrinos, they must be coherent during the production, propagation, and detection processes. In the quantum resource framework and using $l_1\text{-norm}$ as a measure of coherence, this means that the off-diagonal elements of the corresponding density matrix must be non-zero. This happens when the produced neutrino state is a coherent superposition of mass eigenstates, holds it during the propagation and finally is detected coherently. It is noticeable that during the propagation, one can attribute a non-zero entanglement between the various modes of the neutrino flavors which can be evaluated by the usual entanglement measures such as concurrence \cite{bla,mass}. Meanwhile, the coherence of propagation and the entanglement of flavor modes are not independent quantum correlations \cite{mm}. In this section, we will investigate the effects of the charged current elastic forward scattering of electron neutrinos off electrons (matter potential $V$) during the propagation on the $l_1\text{-norm}$. When this quantity is non-zero, we conclude that there exists quantum coherence and one can observe a quantum interference which is the neutrino oscillation.

 One can write the density matrices for the primary electron neutrinos  as
\be
\rho_{e}^m(t)=\begin{bmatrix}
	|A_{ee}^m(t)|^2&A_{ee}^m(t)A_{e\mu}^{m*}(t)&A_{ee}^m(t)A_{e\tau}^{m*}(t)\\
	A_{e\mu}^m(t)A_{ee}^{m*}(t)&|A_{ e\mu}^m(t)|^2&A_{e\mu}^m(t)A_{e\tau}^{m*}(t)\\
		A_{e\tau}^m(t)A_{ee}^{m*}(t)&A_{e\tau}^m(t)A_{e\mu}^{m*}(t)&|A_{ e\tau}^m(t)|^2\\
\end{bmatrix},\label{rhoe}
\ee
and muon neutrinos as
\be
\rho_{\mu}^m(t)=\begin{bmatrix}
	|A_{\mu\mu}^m(t)|^2&A_{\mu\mu}^m(t)A_{\mu e}^{m*}(t)&A_{\mu\mu}^m(t)A_{\mu\tau}^{m*}(t)\\
	A_{\mu e}^m(t)A_{\mu\mu}^{m*}(t)&|A_{\mu e}^m(t)|^2&A_{\mu e}^m(t)A_{\mu\tau}^{m*}(t)\\
	A_{\mu\tau}^m(t)A_{\mu\mu}^{m*}(t)&A_{\mu\tau}^m(t)A_{\mu e}^{m*}(t)&|A_{ \mu\tau}^m(t)|^2\\
\end{bmatrix},\label{rhom}
\ee
where the superscript $m$ emphases that the propagation takes place in the material medium. According to Eq. (\ref{l1}), the $l_1\text{-norm}$s corresponding to the density matrices given in Eqs. (\ref{rhoe}) and (\ref{rhom}) in terms of the oscillation probabilities are
\be
\textit{c}(\rho_e^m)=2(\sqrt{P_{ee}^mP_{e\mu}^m}+\sqrt{P_{ee}^mP_{e\tau}^m}+\sqrt{P_{e\mu}^mP_{e\tau}^m}),
\ee
and 
\be
\textit{c}(\rho_\mu^m)=2(\sqrt{P_{\mu\mu}^mP_{\mu e}^m}+\sqrt{P_{\mu\mu}^mP_{\mu\tau}^m}+\sqrt{P_{\mu e}^mP_{\mu\tau}^m}),
\ee
respectively.  In general, the maximum possible value for $c(\rho)$ is $c_{\text max} = d-1$, where $d$ is the dimension of the corresponding density matrix \cite{plenio2}. Thus it is equal to two here.

\begin{figure}[ht]
	\centering
	\subfigure[]{\includegraphics[scale=0.5]{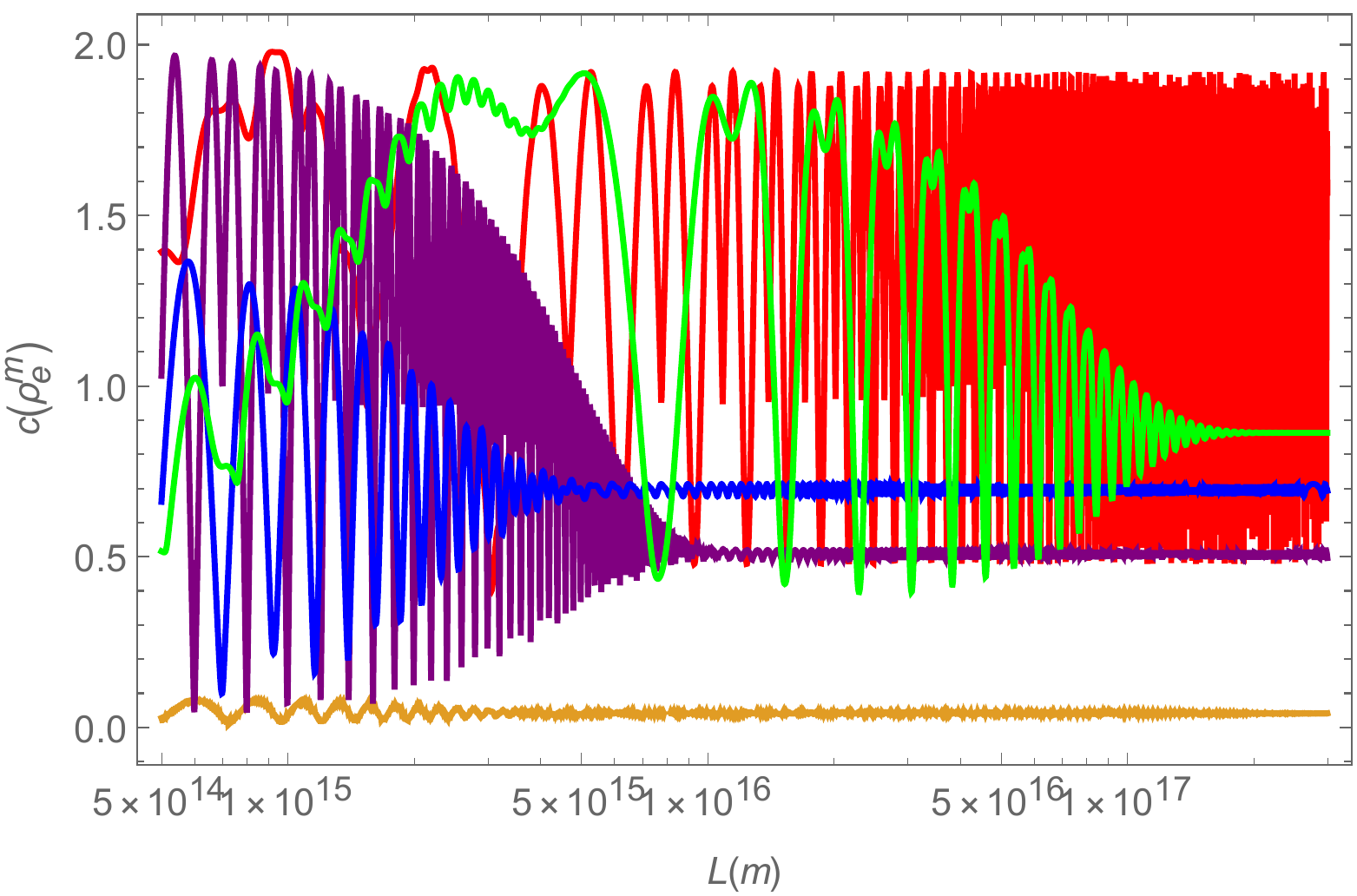}}
	\subfigure[]{\includegraphics[scale=0.5]{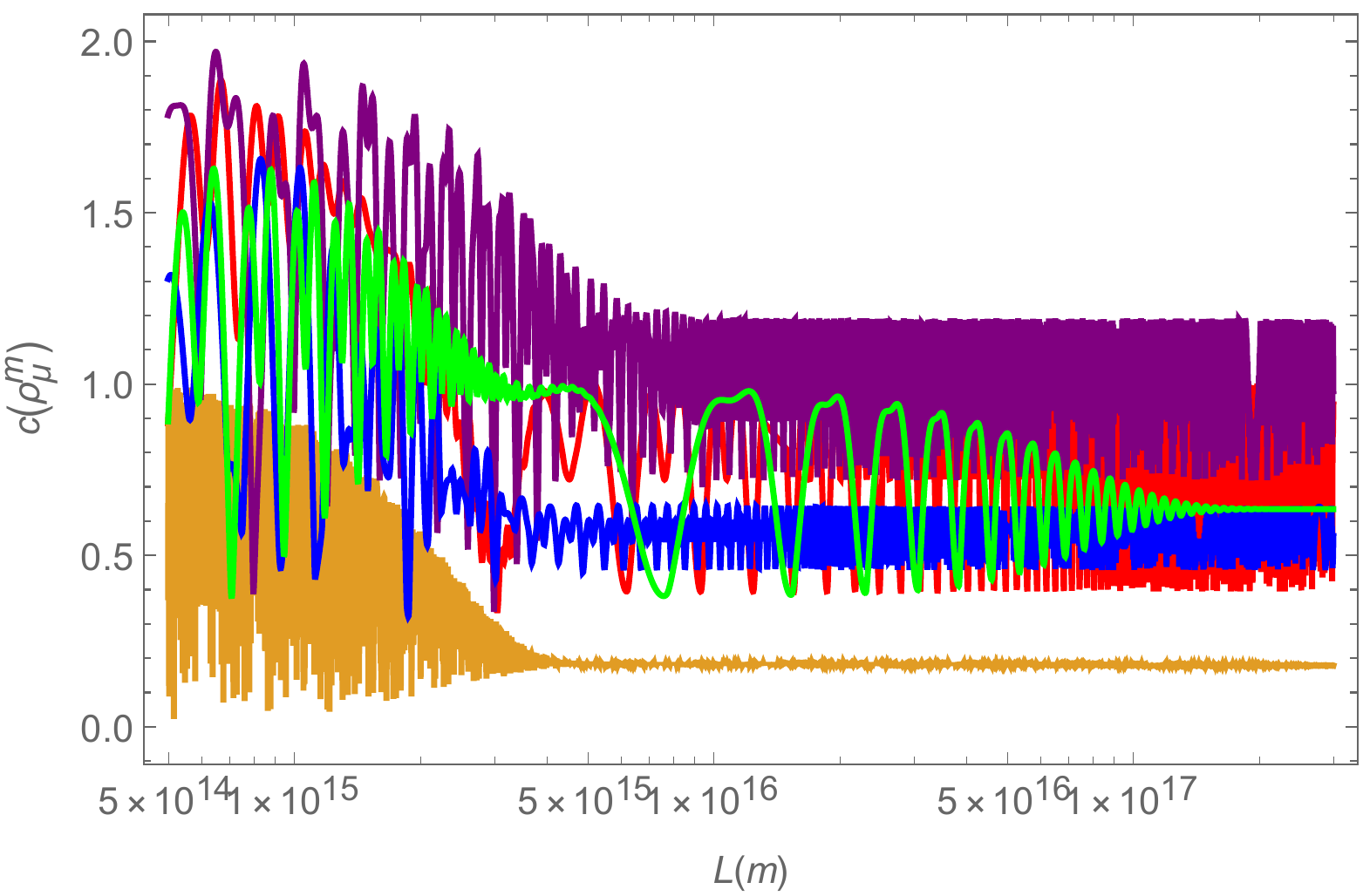}}
	\caption{$l_1\text{-norm}$ for the initial neutrinos to be $\nu_e$ (a) and $\nu_\mu$ (b). In addition to vacuum (green), we take $V=2.242 \times10^{-15} eV$ (red), $V=1.099 \times10^{-14} eV$(blue) and $V=2.824\times10^{-14} eV$(purple) which correspond to infinity coherence lengths. We also take $V=1.00\times10^{-12} eV$(brown) corresponding to matter dominated.}\label{f3}
	
\end{figure}

We illustrate the behavior of $\textit{c}(\rho_e^m)$ (a) and $\textit{c}(\rho_\mu^m)$ (b) versus the baseline distance $L$ through Figs. (\ref{f3}).  Similar to the oscillation probability's illustrations in Figs. (\ref{f1}) and (\ref{f2}), we do these for five cases; vacuum ($V=0$), for three values of potential which correspond to three infinity coherence lengths and for a matter dominated potential. We see that when the initial neutrinos are $\nu_e$ (diagram a in Fig. (\ref{f3})), the $l_1$-norm behavior shows that the coherence condition is conserved if $V=2.242 \times10^{-15} eV$ (the red curve) which cause  to ${L^m}^{\text{coh}}_{21}$ become infinite. It is reasonable because this coherence length is relevant in the transition probabilities. For the other potentials considered in this diagram, the coherence lengths are smaller than vacuum one. When the initial neutrinos are $\nu_\mu$, the $l_1$-norm related to $V=2.824\times10^{-14} eV$ (the purple curve), which leads ${L^m}^{\text{coh}}_{32}$ to become infinite, is larger than the others. Of course, this coherence length is relevant for the probability of transition to $\nu_\tau$. $l_1$-norms for the other cases are smaller than the vacuum one. This shows that except for the two cases mentioned, decoherence in matter is larger than vacuum.

\begin{figure}[ht]
	\centering
	\subfigure[]{\includegraphics[scale=0.5]{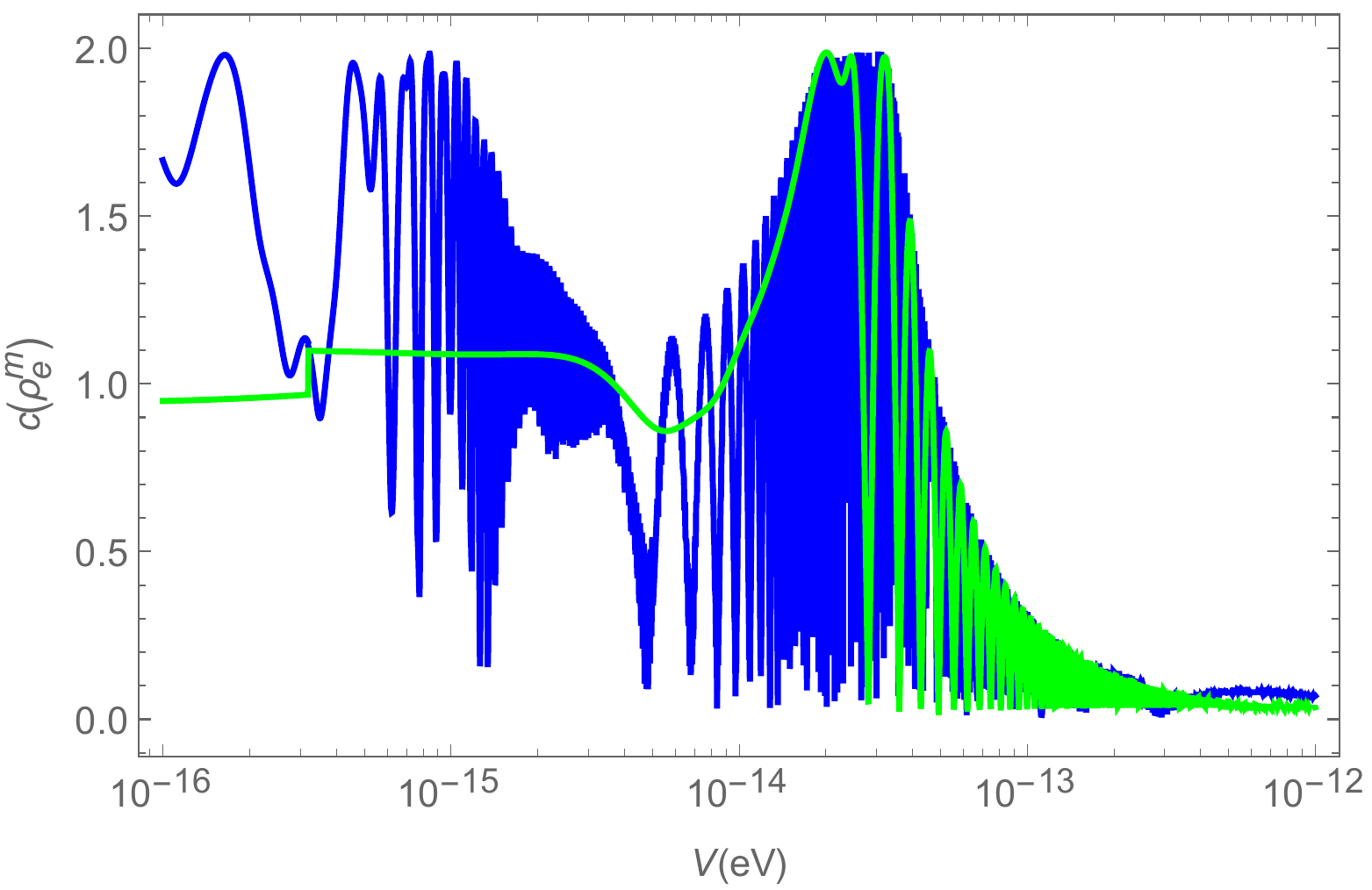}}
	\subfigure[]{\includegraphics[scale=0.5]{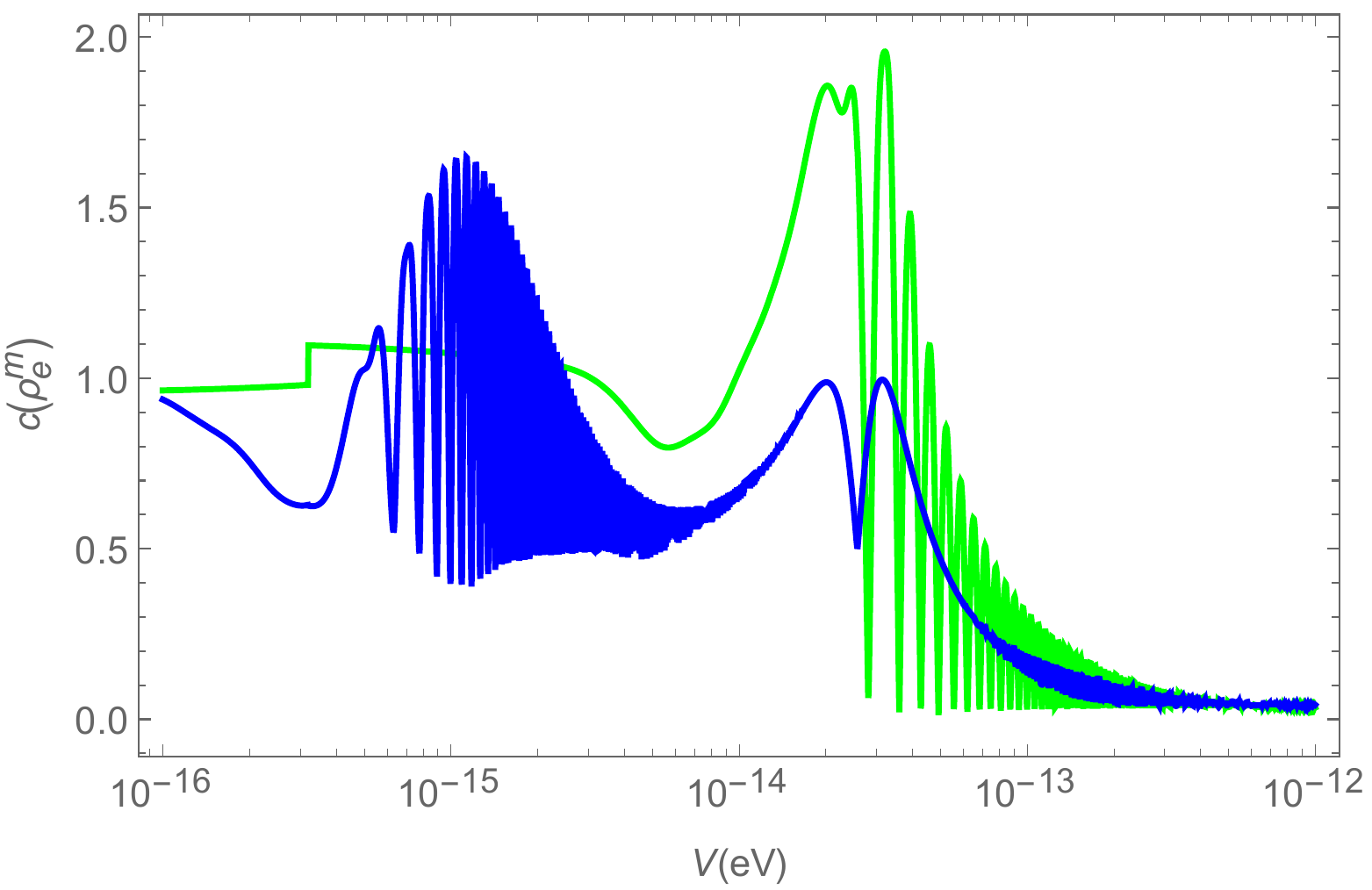}}
	\caption{Plots of $\textit{c}(\rho_e^m)$ against the potential $V$ obtained by PW (a) and WP (b) approach. We take $L=10^{15}m$(green) and $L=10^{17}m$(blue).}\label{f4} 
\end{figure}

\begin{figure}[ht]
	\centering
	\subfigure[]{\includegraphics[scale=0.5]{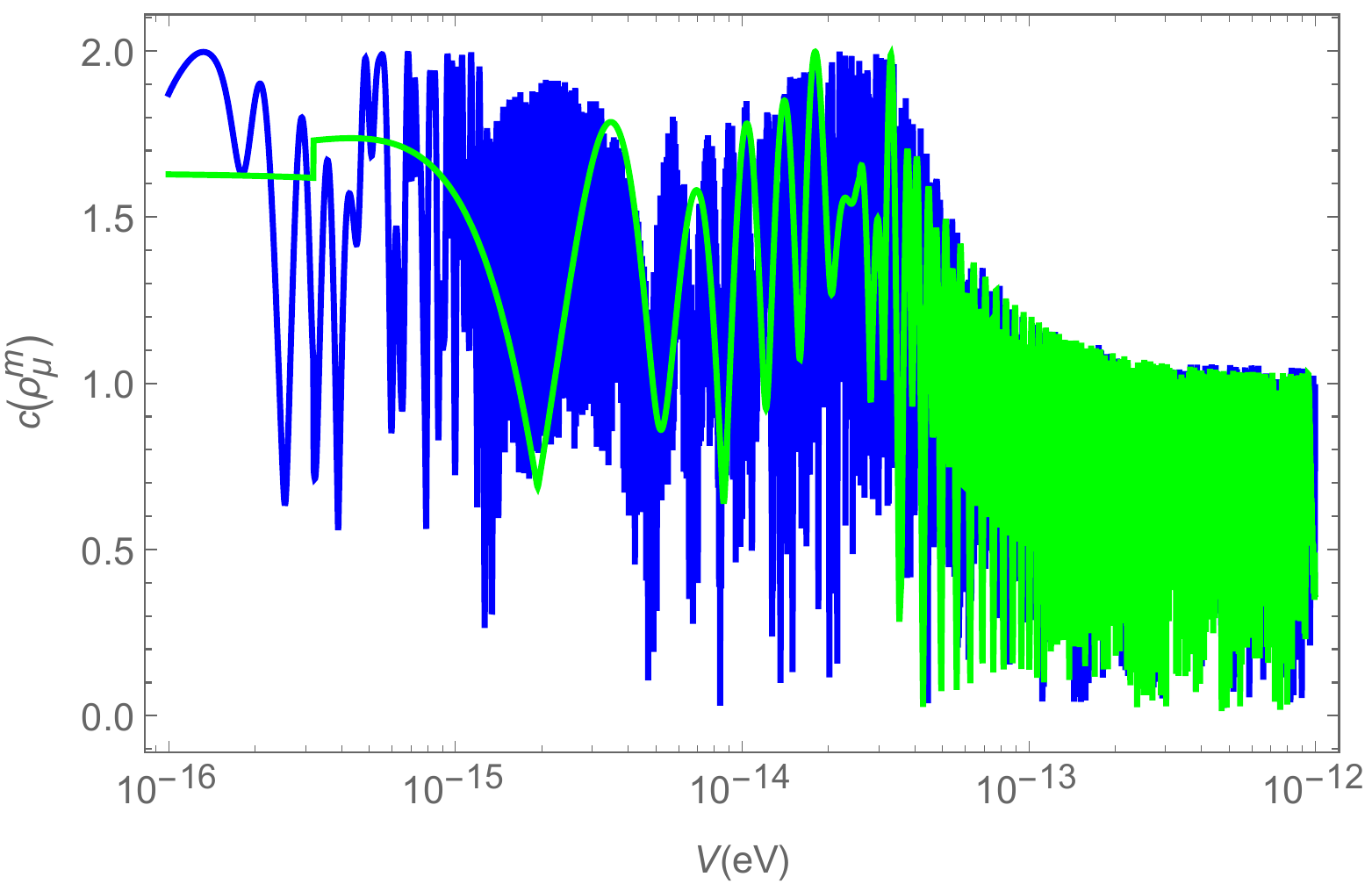}}
	\subfigure[]{\includegraphics[scale=0.5]{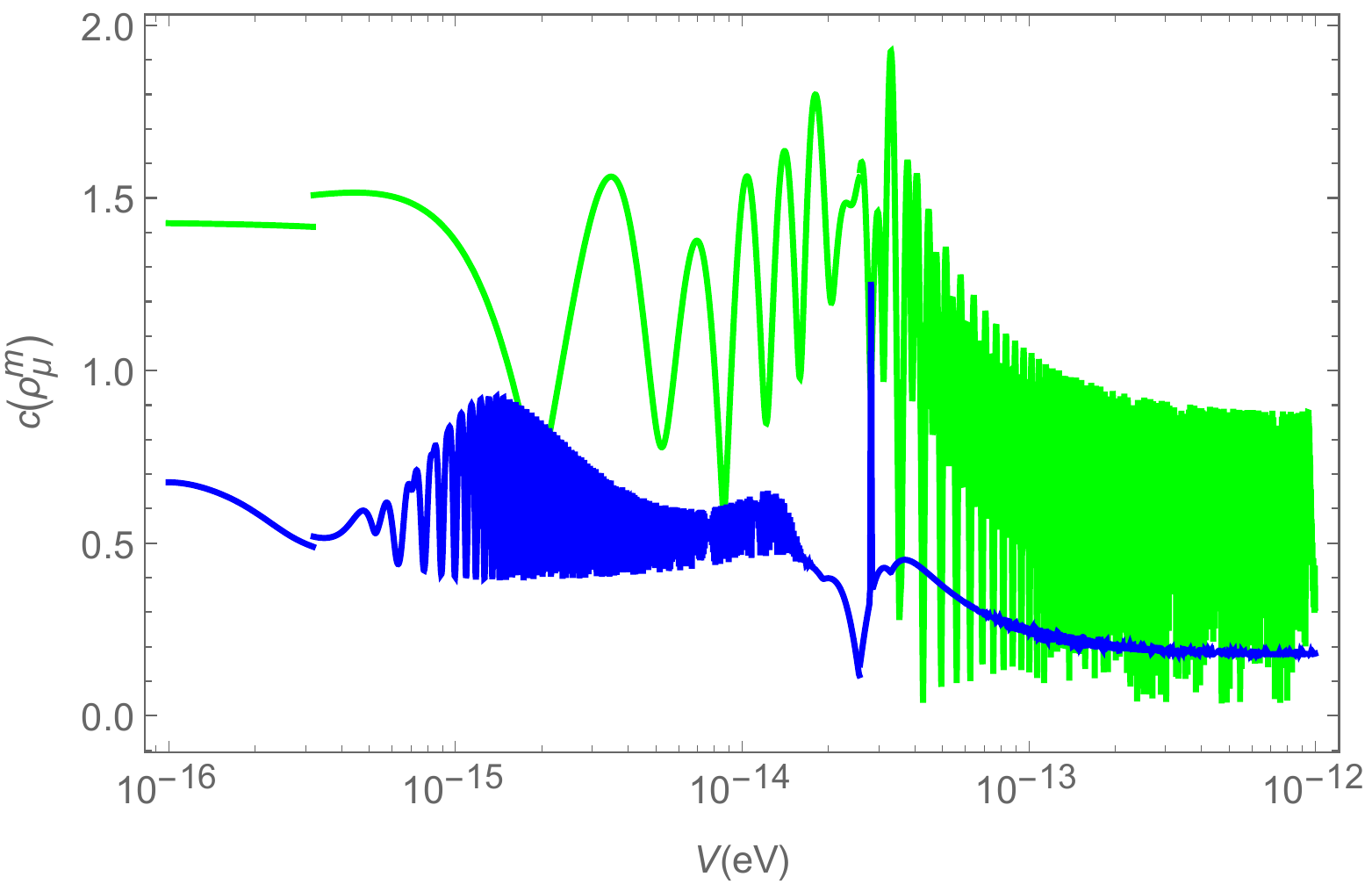}}
	\caption{Plots of $\textit{c}(\rho_\mu^m)$ against the potential $V$ obtained by PW (a) and WP (b) approach. We take $L=10^{15}m$(green) and $L=10^{17}m$(blue).}\label{f5} 
\end{figure}

Furthermore, we compare the results obtained using PW (a) and WP (b) approaches by Figs. (\ref{f4}) and (\ref{f5}). In Figs. (\ref{f4}), $\textit{c}(\rho_e^m)$ is illustrated versus $V$ for two distances $L=10^{15}m$ (green curve) and $L=10^{17}m$ (blue curve), using both PW and WP approaches by the diagrams a and b, respectively.  For the former distance (green curve) which is shorter than the coherence length, $\textit{c}(\rho_e^m)$ obtained by both approaches have similar behavior. In the case of the latter distance, $\textit{c}(\rho_e^m)$ obtained using PW approach achieves the maximum values for two regions of the potential values. However, in the case of WP approach, $\textit{c}(\rho_e^m)$ achieves a maximum smaller than 2 in a region of the potential values. This is due to the decoherence effects coming from the separation of neutrino WPs. We also give a similar illustration for $\textit{c}(\rho_\mu^m)$ in Fig. (\ref{f5}). The statement in this Fig. is similar to previous one.

 Overall, we see that when there exists a perfect decoherence due to the wave packets separation, as well as due to the matter dominated potential, the $l_1$-norm tends to an almost constant value smaller than 1.  Therefore, the other important point resulting from this study is that the $L$-independence of $l_1$-norm can be mooted as an indication of the perfect coherency loss.


\section{Summary and discussions}
Neutrino oscillation is a complex and astonishing quantum phenomenon that researchers have been studying for more than 60 years. So far, after many experiments, we could get a lot of information in this regard, but many aspects are still unknown such as neutrino mass ordering, leptonic CP violation phase and so on. Furthermore, neutrino experiment data can also be utilized as a diverse platform to study the fundamental aspects of quantum mechanics, for instance, see \cite{2016}. In particular, neutrino oscillation can be considered as a good illustration of quantum coherence \cite{song}. There exist a fundamental reason for  the loss of coherency in neutrino oscillation; taking into account the localization properties of production and detection processes, one can find the decoherence effects due to the separation of WPs describing neutrinos \cite{par,kayser,giunti,ettplb,ettscripta}. 
In this paper, in addition to considering this issue, we have investigated the effect of neutrino propagation in the material medium with a constant density on quantum coherence. For this purpose, we  have used the approach introduced in Ref. \cite{1} in order to diagonalize the related Hamiltonian. We know that in addition to neutral current interactions for all neutrinos, the electron neutrino has an additional charge current interaction. Therefore, it is expected that propagation in the material medium leads to a decoherency, which depends on the density of matter in the medium. In fact, this efficient effect is due to the elastic forward scattering of neutrinos off electrons, which causes the electron neutrinos to feel an interaction potential $V$, which is proportional to the density of matter, more than others.  For two values of $V$, we have resonance in the transition processes. However, the coherence length for some values of potentials becomes infinite \cite{Holanda,Kersten,smir2021}. This effect is due to the larger mixing angle in the material medium.

We have illustrated the dependence of the oscillation probabilities and $l_1\text{-norm}$ as a measure for quantum coherence on $V$ through Figs. (\ref{f1}), (\ref{f2}), (\ref{f3}), (\ref{f4}), and (\ref{f5}). We see that the perfect coherence loss due to the wave packets separation, as well as due to the matter dominated potential, causes the $l_1$-norm to tend to almost constant value smaller than 1. In fact, the $L$-independence of $l_1$-norm can be interpreted as an indication of the perfect coherency loss. In general, interaction with matter leads to the faster damping in neutrino oscillation probabilities comparison to vacuum except two values\footnote{We should notice that there are two values for potential ($V=2.242\times 10^{-15}eV$ and $V=1.099\times 10^{-14}eV$) in which ${L^m}^{\text{coh}}_{21}$ becomes infinite. However, in the case of larger value, the decoherence due to the interaction with matter cause this effect to be irrelevant.} of potential in which the coherence lengths ${L^m}^{\text{coh}}_{21}$ and ${L^m}^{\text{coh}}_{32}$ become infinite. 
When the primary neutrinos are electron neutrinos $\nu_e$, ${L^m}^{\text{coh}}_{21}$ plays a significant role in the transition process. Therefore, provided that $V$ has the value in which this coherence length become infinite, the maximum value of $l_1$-norm , $\textit{c}(\rho_e^m)$, is conserved for the interval distance very larger than the coherence length in vacuum (please see Fig. (\ref{f3}) plot a). Also, when the primary neutrinos are muon neutrinos $\nu_\mu$, we have roughly a similar situation for $\textit{c}(\rho_\mu^m)$ (please see Fig. (\ref{f3}) plot b). Furthermore, Figs. (\ref{f4}), and (\ref{f5}) show that $l_1$-norm is naturally independent of the decoherency due to the localization properties when the baseline distance is smaller than coherence lengths. Meanwhile, when the baseline distance is of the order of coherence length, the behaviors of $l_1$-norms obtained by PW approach are nontrivially different from the corresponding one obtained by WP approach. In particular, we find that $\textit{c}(\rho_e^m)$ obtained by PW description has two maximum corresponding to two resonance potentials while the second maximum occurs weakly in the case of WP treatment. 

\appendix
\section{A method for diagonalizing neutrino Hamiltonian in uniform matter }\label{2222}
The scattering of neutrino off electrons in the matter causes the electron neutrinos to feel an additional potential $V$ which varies with position, consequently $t$,  when the density of matter varies with the position. In general, $V(x)$ leads the dispersion relation between the momentum $|\vec{p}|$ and the energy $E$ of neutrinos to be changed. Therefore, the investigation of time evolution of the realistic three generations scheme becomes complicated,
generally speaking. Many authors have considered this problem \cite{paul,joh,wil,smir,niro,ohl,his,shu,akh21,1,2,3}. In particular, neutrino oscillation in the matter has been treated comprehensively in Refs. \cite{2,niro}. More simplified and accurate arguments have been performed for the oscillations in uniform matter density versus baseline divided by neutrino energy plane by using a perturbative framework in Ref. \cite{1}.
As was said, we use the method developed in Ref. \cite{1} to diagonalize the Hamiltonian for a uniform matter. Here we give a brief of it. 



We suppose the flavor eigenstates are written in terms of mass eigenstates as follows:
\be
\left|  \nu _\alpha \right\rangle = \sum_{i=1}^3 U^*_{\alpha i}\left|  \nu _i\right\rangle, \label{flavor}
\ee
where $U_{\alpha i}$ is giving in Eq. \ref{pmns}. One can decompose $U_{\text{PMNS}}$ as follows:
\be
U_{\text{PMNS}}=U_{23}(\theta_{23})I_\delta U_{13}(\theta_{13})I_\delta^*U_{12}(\theta_{12}),
\ee
in which
\be
U_{23}(\theta_{23})=\begin{bmatrix}
	1 & 0 &0 \\
	0&c_{23} &s_{23} \\
	0&-s_{23} &c_{23}  \\
\end{bmatrix},\,\,\,\,\,
I_\delta =\begin{bmatrix}
	1 &0&0 \\
	0&1 &0 \\
	0&0 & e^{i\delta} \\
\end{bmatrix},\,\,\,\,\,
U_{13}(\theta_{13})=\begin{bmatrix}
	c_{13} &0&s_{13} \\
	0&1  &0 \\
	-s_{13}&0&c_{13} \\
\end{bmatrix},\,\,\,\,\,
U_{12}(\theta_{12})=\begin{bmatrix}
	c_{12} & s_{12} &0 \\
	-s_{12}&c_{12}  &0 \\
	0&0 &1  \\
\end{bmatrix}.
\ee
The time evolution of the neutrino state in the presence of matter potential is governed by the following Schrodinger equation:
\be	i\partial_t|\nu_\alpha,t\rangle=\frac{1}{2E}[U_{\text{PMNS}}{\text{diag}}(0,\Delta m_{21}^2,\Delta m_{31}^2)U_{\text{PMNS}}^\dagger+{\text{diag}(2EV(x),0,0)}]|\nu_\alpha,t\rangle,\label{scherodinger}
\ee 
where $V(x)=\sqrt{2}G_FN_e(x)$. $N_e$ denotes the density of electron in the medium and hereafter we assume it to be constant. We know that the neutrino interactions with matter in the medium do not alter CP phase $\delta$ and mixing angle $\theta_{23}$. So Eq. (\ref{scherodinger}) is covariant under the inverse transformation
$U^{-1}_{23}(\theta_{23})$ and $I^{-1}_{\delta}$. Therefore, we can write the Hamiltonian as follows
\be
\acute{H}=\frac{1}{2E}[U_{13}(\theta_{13})U_{12}(\theta_{12}){\text{diag}}(0,\Delta m_{21}^2,\Delta m_{31}^2)U_{12}^\dagger(\theta_{12}) U_{13}^\dagger(\theta_{13}) + {\text{diag}(2EV,0,0)}].
\ee

Now, one can diagonalize $\acute{H}$ by applying two consecutive rotations in the matter.
First, one needs to do the $\theta_{13}^m$ rotation:
\be
H'=U_{13}^{\dagger}(\theta_{13}^m)\acute{H}U_{13}(\theta_{13}^m),
\ee
where
\be
U_{13}^{\dagger}(\theta_{13}^m)=\begin{bmatrix}
	\cos\theta_{13}^m &0&-\sin\theta_{13}^m \\
	0&1  &0 \\
	\sin\theta_{13}^m&0&\cos\theta_{13}^m \\
\end{bmatrix},
\ee
with
\be
\theta_{13}^m=\dfrac{1}{2}\arctan[\dfrac{\epsilon\sin{2\theta_{13}}}{\epsilon\cos{2\theta_{13}}-2EV}].
\ee
Here, $\epsilon$ is given by
\be
\epsilon=\Delta m_{31}^2-\Delta m_{21}^2\sin^2{\theta_{12}}.
\ee
Therefore, $H'$ is obtained as follows:
\bea{\nonumber}
H'&=&\dfrac{1}{2E}\begin{bmatrix}
	l_1&0&0\\
	0&l_2&0\\
	0&0&l_3\\
\end{bmatrix}\\
&+&\dfrac{\Delta m_{21}^2\sin{2\theta_{12}}}{4E}\begin{bmatrix}
	0&\cos(\theta_{13}-\theta_{13}^m)&0\\
	\cos(\theta_{13}-\theta_{13}^m)&0&-\sin(\theta_{13}-\theta_{13}^m)\\
	0&-\sin(\theta_{13}-\theta_{13}^m)&0\\
\end{bmatrix},
\eea
in which
\be
l_2=\Delta m_{21}^2\cos^2{\theta_{12}},
\ee
\be
l_{1,3}=\dfrac{1}{2}[(\Delta m_{31}^2+2EV+\Delta m^2_{21}\sin^2\theta_{12})\mp\sqrt{(2EV)^2+\epsilon^2-4EV\epsilon \cos{2\theta_{13}}}].
\ee
Second, a rotation with $\theta_{12}^m$ angle must be performed such a way that $H'$ is transformed as follows:
\be
H''=U_{12}^{\dagger}(\theta_{12}^m)H'U_{12}(\theta_{12}^m)=H_0+H_1,\label{15}
\ee
with
\be
H_0=\begin{bmatrix}
	E^m_1&0&0\\
	0&E^m_2&0\\
	0&0&E^m_3\\
\end{bmatrix},
\ee
where
\be
\label{E3}
E^m_3=\frac{l_3}{2E},
\ee
\be
\label{E12}
E^m_{1,2}=\frac{1}{4E}[(l_1+l_2)\mp\sqrt{(l_1-l_2)^2+\Delta m_{21}^4\sin^2{2\theta_{12}}\cos^2({\theta_{13}}-\theta_{13}^m})],
\ee
and
\be
H_1=\dfrac{1}{4E}\Delta m_{21}^2 \sin{2\theta_{12}}\sin(\theta_{13}-\theta_{13}^m)\begin{bmatrix}
	0&0&\sin{\theta_{12}^m}\\
	0&0&-\cos{\theta_{12}^m}\\
	\sin{\theta_{12}^m}&-\cos{\theta_{12}^m}&0\\
\end{bmatrix}.
\ee
$U_{12}^{\dagger}(\theta_{12}^m)$ is defined by
\be
U_{12}^{\dagger}(\theta_{12}^m)=\begin{bmatrix}
	\cos\theta_{12}^m & -\sin\theta_{12}^m &0 \\
	\sin\theta_{12}^m&\cos\theta_{12}^m  &0 \\
	0&0 &1  \\
\end{bmatrix},
\ee
in which
\be
\theta_{12}^m=\dfrac{1}{2}\arctan[\Delta m_{21}^2\dfrac{\sin{2\theta_{12}}}{l_2-l_1}\cos(\theta_{13}-\theta_{13}^m)].
\ee

In Eq. (\ref{15}), $H_0$ has the main contribution in the total Hamiltonian $H''$ and one can treat perturbatively with $H_1$. However, in this paper we ignore the $H_1$ contribution because it will have no remarkable effects on our results.


In the case of anti-neutrino, the following changes are necessary:
\be
I_\delta\longrightarrow I_\delta^*,
\ee
and
\be 
V\longrightarrow -V.
\ee 
Therefore, the final parameters of the anti-neutrino oscillation in matter are given by
\be
\bar{\theta}_{13}^m=\dfrac{1}{2}\arctan[\dfrac{\epsilon\sin{2\theta_{13}}}{\epsilon\cos{2\theta_{13}}+2EV}],
\ee
\be
\bar{\theta}_{12}^m=\dfrac{1}{2}\arctan[\Delta m_{21}^2\cos(\theta_{13}-\bar{\theta}_{13}^m)\dfrac{\sin{2\theta_{12}}}{\bar l_2-\bar l_1}],
\ee

\be\label{8}
{\bar E}^m_3=\frac{\bar l_3}{2E},
\ee
\be
\label{9}
{\bar E}^m_{1,2}=\dfrac{1}{4E}[(\bar l_1+\bar l_2)\mp\sqrt{(\bar l_1-\bar l_2)^2+\Delta m_{21}^4\sin^2{2\theta_{12}}\cos^2({\theta_{13}}-\bar\theta_{13}^m})],
\ee
with
\be
\bar l_2=l_2=\Delta m_{21}^2\cos^2{\theta_{12}},
\ee
\be
\bar l_{1,3}=\dfrac{1}{2}[(\Delta m_{31}^2-2EV+\Delta m_{21}^2\sin^2{\theta_{12}})\mp\sqrt{(2EV)^2+\epsilon^2+4EV\epsilon \cos{2\theta_{13}}}].
\ee




\end{document}